# Enhanced light-vapor interactions and all optical switching in a chip scale micro-ring resonator coupled with atomic vapor

Liron Stern[1], Roy Zektzer[1], Noa Mazurski[1], and Uriel Levy[1,2]

[1]Department of Applied Physics, The Benin School of Engineering and Computer Science, The Center for Nanoscience and Nanotechnology, The Hebrew University of Jerusalem, Jerusalem, 91904, Israel

[2]ulevy@mail.huji.ac.il

**Abstract:** The coupling of atomic and photonic resonances serves as an important tool for enhancing light-matter interactions and enables the observation of multitude of fascinating and fundamental phenomena. Here, by exploiting the platform of atomic-cladding wave guides, we experimentally demonstrate the resonant coupling of rubidium vapor and an atomic cladding micro ring resonator. Specifically, we observed cavity-atom coupling in the form of Fano resonances having a distinct dependency on the relative frequency detuning between the photonic and the atomic resonances. Moreover, we were able to significantly enhance the efficiency of all optical switching in the V-type pump-probe scheme. The coupled system of micro-ring resonator and atomic vapor is a promising building block for a variety of light vapor experiments, as it offers a very small footprint, high degree of integration and extremely strong confinement of light and vapor. As such it may be used for important applications, such as all optical switching, dispersion engineering (e.g. slow and fast light) and metrology, as well as for the observation of important effects such as strong coupling, Purcell enhancement and bistability.

In recent years we are witnessing a flourish in research aimed to facilitate alkali vapors in guided wave configurations[1–16]. Owing to the significant reduction in device dimensions, the increase in density of states, the interaction with surfaces and primarily the high intensities carried along the structure, a rich world of light vapor interactions can be studied, and new functionalities such as low power nonlinear light-matter interactions can be achieved. Further enhancement of light matter interactions can be obtained by implementing guided-light and vapor interactions in resonator systems such as Fabry-perot cavities and micro ring resonators (MRRs)[4,17]. Indeed, the combination of photonic and atomic resonances are of fundamental and applicative interest as a myriad of fascinating phenomena such as strong coupling[18], Purcell enhancements[19], Fano resonances[13,20], slow and fast light enhancement[21], and cavity broadening[22] can now be investigated.

Here, we report on the experimental realization of a coupled micro-ring and atomic system which we defined as an atomic cladding micro-ring resonator (ACMRR). By integrating an atomic vapor with a MRR, we were able to witness clear signatures of the coupling between the atomic and photonic modes manifested as Fano resonances. Such Fano resonances are controlled by varying the detuning between the MRR resonance and the atoms via the thermo optic effect. Furthermore, we excite the coupled system with a pump and probe operating in the V-type scheme. We observe a significant reduction of the optical power of the pump beam needed for inducing the transparency window. This enhancement is correlated to the detuning of the pump beam with respect to the MRR resonance. This enhancement is a direct consequence of the buildup of electro-magnetic energy within the MRR. Such ACMRRs may utilized in a variety applications spanning from all optical switching and slow light to the observation of strong coupling and Purcell enhancements.

In Fig. 1a we present an artistic rendering of our coupled ACMRR system. The system consists of a photonic chip with a 600nm width by 250 nm height silicon nitride (SiN) waveguide coupled to a MRR. The entire chip is covered with a 2-micron thick silicon dioxide cladding, except for a specific region around the MRR where the oxide layer is removed by wet etching. This opening will later be used as the interaction region between the MRR and the Rb atoms. In all other areas of the chip, expect this interaction region, the waveguides are buried under the oxide layer. Next, a Pyrex cylinder is epoxy bonded to the chip and subsequently connected to a vacuum system, baked out, and a droplet of $^{85}$Rb is inserted. The fabrication method of such an integrated atomic cladding wave guide is described in detail in ref. 14. The waveguide cross section with the superimposed electric field norm of the fundamental mode transverse electric like mode is sketched in Fig. 1b.

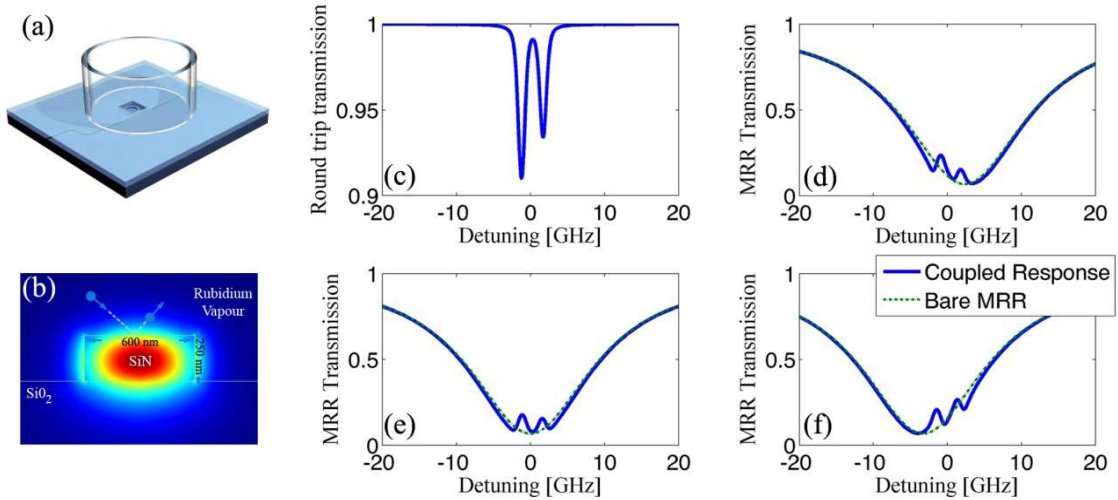

**Figure 1**: **(a)** Schematic illustration of an atomic vapor cell bonded to a chip consisting of a SiN MRR **(b)** A sketch showing the waveguide cross section with the superimposed electric field norm of the fundamental mode transverse electric like mode. **c)** Round trip transmission (normalized linear scale) of the atomic cladding wave guide which constitutes the MRR **(d)** ACMRR response for an MRR which is blue-detuned from the atomic resonance (blue line) **(e)** ACMRR response in for an MRR which is tuned to the atomic resonance (blue line) **(f)** ACMRR response in for an MRR which is red-detuned from the atomic resonance (blue line). The green curves in panels d-f correspond to the MRR resonance in absence of Rb medium.

Next, we describe the expected transmission of the coupled ACMRR system. To do so, we use the model presented in ref. [21] to calculate the transmission of such system. In short, we adapt the formalism of attenuated total internal reflection[23–25] to account for our configuration. First, we find the effective susceptibility of the atomic media serving as a top cladding, taking into account the motion of the atoms, their quenching on the surface and the wave vectors of the evanescent waveguide mode. As such, it includes two primary broadening mechanisms: Enhanced (relative to vacuum) Doppler broadening, and transit time broadening. A detailed description of the procedure is given in ref. 14, and has been found to yield good results in a variety of guided wave configurations[7,13,14,26]. Next, we use this effective susceptibility to represent the top cladding of our MRR, using the well known transfer functions of an MRR, and taking into account the confinement ratio of the optical mode interacting with the Rb vapor. In Fig. 1c we present the calculated normalized round trip transmission of the atomic cladding wave-guide constituting the MRR (calculated by assuming a waveguide length which corresponds to the MRRs circumference with Rb density of $5 \times 10^{12}$ cm$^{-3}$). The MRR radius is assumed to be 80µm, its Q factor was chosen to be moderate (~20,000), and the MRR was assumed to be in the over coupling regime, matching our experimental results to be reported later. Following, using this formalism we plot (Fig. 1d-1f) the ACMRRs response for three different frequency detuning values between the bare MRR (i.e. an MRR with vacuum as its top cladding) and the Rb atomic lines. In Fig 1d the ACMRR response in the case where the MRR is blue detuned from the atomic lines is presented. As can be observed, the response of the combined system exhibits a Fano lineshape[20]. Such fundamental phenomenon is the typical response of systems where a broad resonance is coupled to a narrow line. In the case described here, the atomic medium serves as the narrow resonance while the MRR as the broad one. Interestingly, one can observe two different types of Fano lineshapes in the same spectrum. This is because the MRR lines are fairly narrow (~20 GHz), i.e. not much broader as compared with the separation between two adjacent Rb lines (3 GHz). This is in contrast to our previously reported results dealing with a coupled plasmonic-atomic system[13] where the ultra-broad plasmonic resonance is orders of magnitude larger than the atomic linewidth. Indeed, the $^{85}$Rb F=3 to F' coupled transition resembles an anomalous dispersion lineshape, whereas the $^{85}$Rb F=2 to F' coupled transition shows a transmission peak. The former corresponds to an atomic line which is red detuned from MRR resonance, whereas the latter corresponds to the scenario of relatively aligned resonances. Subsequently we plot in Fig. 1e the ACMRR response in the case of vanishing detuning between MRR and atomic system. Clearly, the atomic resonances appear as peaks, as anticipated for a hybrid atomic-MRR system in the over coupling regime [21]. Finally, for the red detuned MRR, plotted in Fig. 1f, we observe an opposite scenario with respect to the red detuned one of Fig. 1d. Here, the calculations predict a normal dispersion-like linshape for the F=2 to F' transition, and a peak lineshape for the F=3 to F' line.

Following this discussion, we now turn to describe our experimental results. In Fig. 2a we present a microscope micrograph of our SiN MRR, having a radius of 85µm and ring-waveguide coupling gap of ~170nm. Details regarding the fabrication of such MRRs is available for instance in ref. [27]. Next, we present a typical transmission spectrum of our bare MRR, before introducing the Rb atoms. We observed Q-factor values in the range of few thousands to few tens of thousands, with typical extinction ratios ranging from 2dB to 6dB. In order to align the MRR resonance with the Rb resonance we control the operating temperature of the chip by using resistive heaters via the thermo optic effect. It should be noted that changing the temperature is also affecting the atomic density, which impose a practical limitation on the range of MRR resonance tuning. Nevertheless, as we have several MRRs on the chip, and given the moderate free spectral range (~0.5 nm), it is possible to find MRRs having a resonance line in the vicinity of the D2 Rb line.

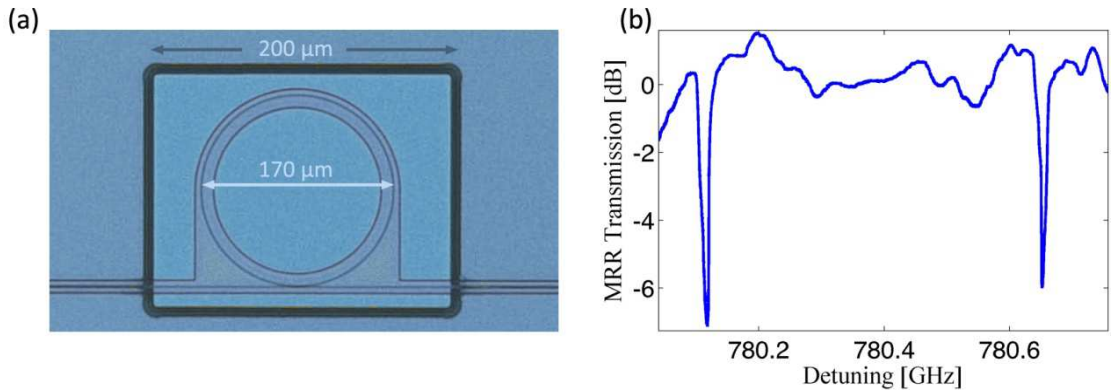

**Figure 2**: (a) A microscope micrograph showing a top view of the MRR (b) Typical measured transmission spectrum of the bare SiN MRRs, before introducing Rb vapor.

Next, we plot our ACMRR response for three different detunings. We plot both the ACMRR transmission (green lines, normalized arbitrarily to the maximal value presented in the figure), the normalized Rb response (blue lines) and natural Rb line reference curves (red dotted lines). In Fig. 3a the F=3 transition is approximately 7.5 Ghz red detuned from the MRR, whereas in Fig. 3b the transition is approximately aligned with the MRR resonance. Finally, in Fig. 3c the transition is ~7 GHz blue detuned from the MRR. Evidently, Fano lineshapes which are highly resembling the calculated spectra (Fig. 1d-1f) are observed, including the observance of two types of Fano resonances in the same spectrum (see e.g. Fig. 3a). Such resonances serve as a clear indication of the hybridization of the atomic and photonic resonance.

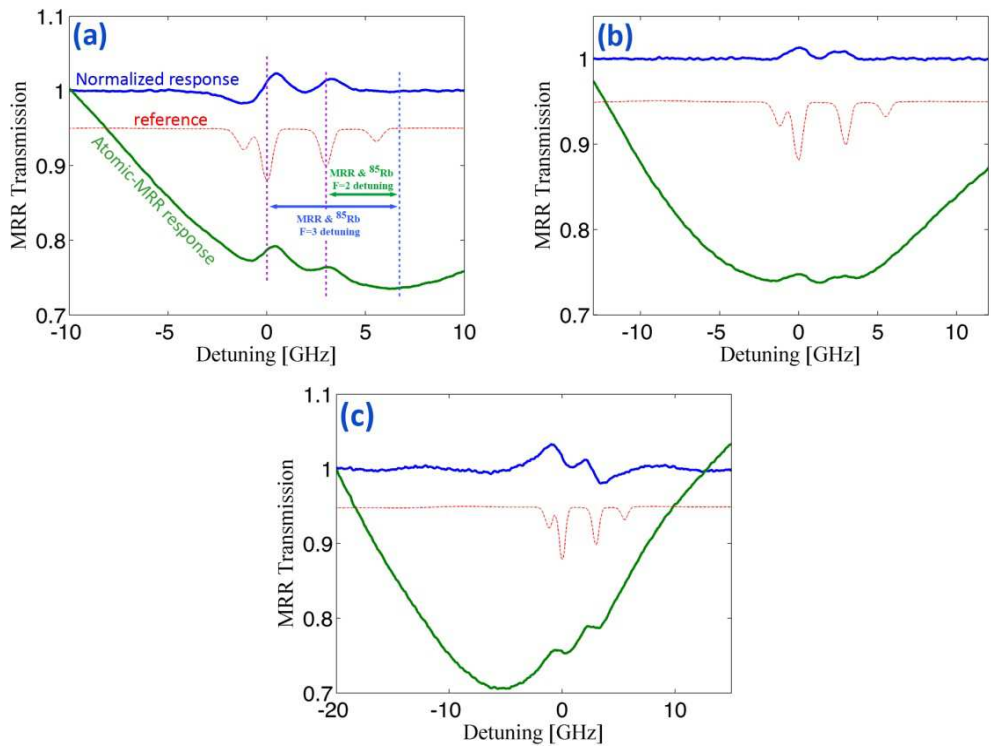

**Figure 3**: ACMRR response for different detunings between the atomic transition and the MRR resonance in the case of **(a)** blue detuned MRR resonance with respect to the atomic transitions **(b)** MRR resonance aligned in frequency with the atomic transitions **(c)** red detuned MRR resonance with respect to the atomic transitions. The green line represents the ACMRR

transmission, the blue line represents the normalized response, achieved by dividing the raw (green line) by a fit to the MRR resonance excluding the Rb contribution, and the red line represents the spectrum of a natural Rb reference cell.

After studying the properties of the hybrid atomic-photonic resonant system and demonstrating the effect of coupling on the lineshapes, we turn into the demonstration of its usefulness for the application of all optical switching. In recent years, great effort is made to facilitate all optical switching, with the ultimate goal of achieving few photon and even single photon switch. Here, we implement the all optical switching by a pump-probe apparatus aimed to modulate the intensity of a probe beam by applying pump beam at low optical powers taking advantage of both confinement and enhancement provided by our ACMRR. In Fig. 4a we sketch the relevant level scheme of $^{85}$Rb levels of the D1 and D2 lines. Operating in the V-type scheme the pump beam at 780nm wavelength is tuned to the F=3 to F' manifold, centered at the middle of the Doppler broadened overlapping transitions. In contrast, the 795nm probe beam is scanned across both ground states (F=2 & F=3) to excited states transitions. We couple both pump and probe in co-propagating manner into our ACMRR, as illustrated in Fig. 4b. In order to isolate the probe beam from the relatively strong pump beam we use band pass filters which eliminate the 780nm pump wavelength. In Fig. 4c, we plot the measured probe spectrum with (blue line) and without (green line) the pump. As illustrated in the inset (Here, blue lines represent the Rb reference, and the violet lines the MRRs spectrum) the MRR is aligned with the atomic resonance. As can be evident, the F=3 absorption is almost totally diminished. The diminished absorption is attributed to the combination of the Autler-Townes effect and the saturation effect as described in ref. [26]. Additionally, the F=2 transition is also influenced by the presence of the pump, as evident by the blue-shifted line. This shift is the result of the AC stark effect, and is estimated from the measured results to be in the order of 200MHz. In Fig. 4d we plot the transmission spectrum of the probe in the case where the MRR resonance is detuned from the pump beam by about 25 GHz, as evident by the position of the blue detuned MRR resonance plotted in the inset. Here the contrast of the optical modulation of the F=3 transition is significantly weakened, as evident by the reduced change in absorption. Furthermore, the light shift is also absent. By comparing the results of Fig. 4c and 4d it is evident that the all optical modulation effect depends on the detuning between the MRR resonance and the pump. This can be simply explained by the buildup of the pump field within the MRR. This buildup is maximal around resonance, and falls off rapidly with the detuning. Interestingly, although the probe beam is not perfectly aligned with one of the MRR resonances, there is still a sufficient field overlap to achieve modulation enhancement.

As a final point we estimate the power levels within the waveguide and the MRR. For the pump beam, we estimate the power in the waveguide to be ~$3\mu W$. In a previous report[26], we have estimated the switching power levels to induce a full turn off of the transition to be ~$14\ \mu W$. Thus we can deduce that we have a five-fold power reduction for all optical switching applications in respect to a bare wave guide configuration. This conclusion is also supported by observing the light shift. Indeed, the 200 MHz observed from Fig. 4c agrees with the light shift observed in our previous paper, corresponding to ~$14\ \mu W$. The switching power can be further reduced simply by improving the Finesse of the resonator. As Finesse values approaching $10^4$ have been observed in such SiN based resonators[28], we believe that nW switching power should be feasible. Furthermore, one can consider other cavities, e.g. micro toroids[29], in which the Finesse approaches spectacular values around $10^6$. With such resonators, one could expect switching power of ~10 pW, corresponding to the few photon and even the single photon regime, albeit with limited switching speed due to the large photon lifetime within the resonator.

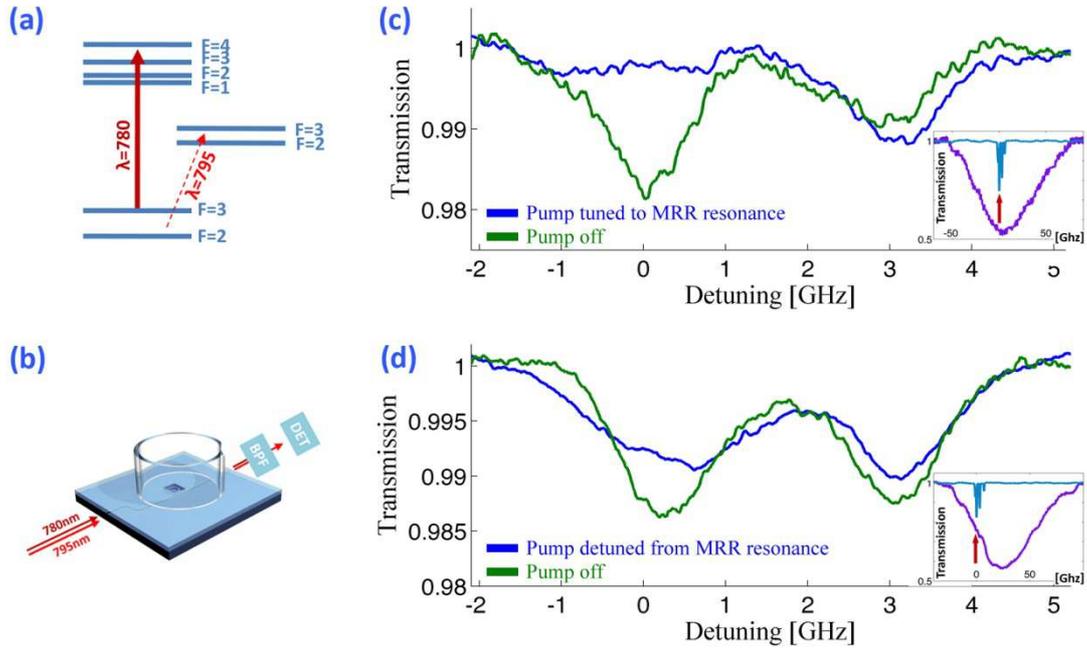

**Figure 4: (a)** Relevant $^{85}$Rb level diagram for the V-type scheme **(b)** Schematic illustration of the all optical switching setup (BPF - band pass filter, DET - detector) **(c)** Probe (795nm) transmission spectra in the presence (blue line) and the absence (green line) of a pump beam (780nm) tuned to the $^{85}$Rb F=3 to F' 2/3/4 manifold and to the MRRs resonance as illustrated in the inset (Natural Rb spectrum represented by the blue line and the MRR is represented by the violet line) **(d)** Probe (795nm) transmission spectra in the presence (blue line) and the absence (green line) of a pump beam (780nm) tuned to the $^{85}$Rb F=3 to F' 2/3/4 manifold and detuned from the MRRs resonance ($\Delta = 24\ GHz$) as illustrated in the inset.

To summarize we have experimentally realized an atomic cladding micro ring resonator (ACMRR). Using this device, we were able to witness clear Fano resonances with distinct dependency on the atomic and MRR frequency detuning. Such spectrum agrees well with a model taking into account the different broadening mechanisms akin to atomic cladding waveguides. Interestingly, in such a system we were able to witness two different types of Fano resonance in the same spectrum. This typecast of resonance is a clear indication of coupling between the Rb and the MRR resonance. Furthermore, we were able to utilize the system to enhance the contrast of the transparency window in a pump-probe all optical switching apparatus. Clear enchantment of the contrast of the transparency window as a function of the pump beam detuning from resonance has been observed. We estimate a five-fold increase in the buildup of intensity within the MRR to be responsible to this enhanced operation. The power needed to switch the probe is estimated to be in the few (~3) microwatt regime, i.e. about 5 times lower as compared with our previously reported result.

The ACMRR is a powerful system to study and implement light-vapor interactions and applications on chip. The interplay between the photonic and the atomic resonances gives rise to rich and diverse set of tools for the control on the different degrees of freedom of light on chip. For example, such a device can be used to control the speed of light, by exploiting the coupling between the group index of the atomic and structural resonances[21]. Furthermore, as the atoms are highly non-linear, a bi-stable operation may be explored. The latter may shown to be useful in memory applications. Further enhancement of the effects demonstrated here, e.g. by increasing the contrast and efficiency of switching can be explored by increasing the atomic density and by improving the Q-factors of the MRRs. The latter may enable the observation of strong coupling in such system.

**Acknowledgments**:

The authors thank Avinoam Stern, Yefim Barash and Benny Levy from AccuBeat Ltd. for the use of the vacuum facilities. We thank Tilman Pfau and Robert Löw for fruitful discussions. The authors acknowledge funding from the ERC grant LIVIN, and the Israel Science Foundation. The waveguides were fabricated at the Center for Nanoscience and Nanotechnology, The Hebrew University of Jerusalem.